\documentclass[sigconf,balance=false]{acmart}

\usepackage{color}
\usepackage{subcaption}
\usepackage{xspace}
\usepackage{multirow}
\usepackage{pifont}
\usepackage[normalem]{ulem}
\usepackage[absolute,overlay]{textpos}

\newcommand{\mt}[1]{{\color{orange} MT: [#1]}}

\copyrightyear{2026}
\acmYear{2026}
\setcopyright{cc}
\setcctype{by}
\acmConference[CCS '26]{Proceedings of the 2026 ACM SIGSAC Conference on Computer and Communications Security}{November 15--19, 2026}{The Hague, Netherlands}
\acmBooktitle{Proceedings of the 2026 ACM SIGSAC Conference on Computer and Communications Security (CCS '26), November 15--19, 2026, The Hague, Netherlands}
\acmDOI{10.1145/3830454.3832633}
\acmISBN{979-8-4007-2871-6/2026/11}

\begin{document}

\title[The Rise and Fall of Google's Privacy Sandbox]{The Rise and Fall of Google's Privacy Sandbox}


\author{Rachid Youssef Grib}
\affiliation{%
  \institution{Politecnico di Torino}
   \city{Torino}
   \country{Italy}
  }
\email{rachidyoussef.grib@studenti.polito.it}

\author{Alberto Verna}
\affiliation{%
  \institution{Politecnico di Torino}
   \city{Torino}
   \country{Italy}
  }
\email{alberto.verna@polito.it}

\author{Nikhil Jha}
\affiliation{%
  \institution{Politecnico di Torino}
   \city{Torino}
   \country{Italy}
  }
\email{nikhil.jha@polito.it}

\author{Martino Trevisan}
\affiliation{%
  \institution{University of Trieste}
   \city{Trieste}
   \country{Italy}
  }
\email{martino.trevisan@dia.units.it}

\author{Marco Mellia}
\affiliation{%
  \institution{Politecnico di Torino}
   \city{Torino}
   \country{Italy}
  }
\email{marco.mellia@polito.it}


\renewcommand{\shortauthors}{Grib et al.}

\begin{abstract}
On October 17\textsuperscript{th}, 2025, Google announced the retirement of most Privacy Sandbox APIs, concluding nearly five years of experimentation with its alternative to privacy-invasive data collection on the Web. Designed to balance privacy with advertising functionality and cross-site tracking, the initiative faced repeated redesigns and limited ecosystem support.

In this work, we present the first longitudinal, consent-aware measurement of the Privacy Sandbox's deployment across the Web. Using a custom call listener and weekly crawls of the top-10,000 websites, we monitor the usage of all major APIs in the months preceding their retirement. Adoption had already stagnated well before Google's announcement: most APIs were used by only a handful of actors, whose activity declined steadily throughout our study. Even the APIs that Google plans to maintain show no sign of growth. The sole exception is Cookies Having Independent Partitioned State (CHIPS). Overall, the demise of the Privacy Sandbox leaves unresolved the challenge of enabling privacy-preserving interest-based advertising.
\end{abstract}

\keywords{Web privacy, Web measurement, Privacy Sandbox}

\TPshowboxestrue
\TPMargin{0.3cm}
\begin{textblock*}{15.5cm}(3cm,1.2cm)
\bf
\centering
\definecolor{myRed}{rgb}{0.55,0,0}
\color{myRed}
\noindent
Please cite this article as: Rachid Youssef Grib, Alberto Verna, Nikhil Jha, Martino Trevisan, Marco Mellia. The Rise and Fall of Google's Privacy Sandbox. ACM CCS 2026. DOI: \href{https://doi.org/10.1145/3830454.3832633}{10.1145/3830454.3832633}
\end{textblock*}

\maketitle

\section{Introduction}
\label{sec:intro}

Online tracking underpins today's digital economy by enabling extensive profiling of users based on their behaviour, interests, and browsing patterns. These profiles fuel the targeted advertising ecosystem and sustain much of the content and services available on the Web~\cite{mayer2012third, wei2020twitter}. However, the traditional mechanisms that support this ecosystem, most notably third-party cookies and browser fingerprinting~\cite{englehardtTracking, roesnerTracking}, enable pervasive cross-site tracking, raising long-standing privacy concerns and attracting increasing regulatory scrutiny.

In response, browser vendors have begun deploying countermeasures, including blocking third-party cookies by default~\cite{firefoxTPCPhaseOut, safariTPCPhaseOut} and proposing privacy-preserving replacements~\cite{googlePrivacySandbox, braveEphemeralTPC, firefoxTotalCookieProtection, appleAdAttribution, firefoxPrivacyAttribution, microsoftNewAdsAPIs}.
These efforts expose a structural tension: while privacy regulation and user expectations increasingly constrain unrestricted tracking, targeted advertising remains a central economic driver of the Web.

Google's Privacy Sandbox represents the most ambitious attempt to resolve this tension through a browser-enforced redesign of advertising and tracking primitives. Rather than exposing stable identifiers or unrestricted browser state (e.g., via persistent third-party cookies), the Privacy Sandbox proposes a collection of APIs that move sensitive processing (e.g., user profile creation, ad selection, and reporting) from remote servers to the user's browser. The browser acts as a trusted enforcement point, locally executing privacy-sensitive logic and only releasing limited, aggregated or noisy signals to third parties on a controlled basis.

Since its first announcement in 2019, the Privacy Sandbox has undergone repeated redesigns, public testing phases, and changes in deployment strategy~\cite{privacySandboxExpanding}. APIs reached general availability between 2023 and 2024~\cite{privacySandboxGeneral}, originally intended to coincide with Chrome’s phase-out of third-party cookies~\cite{googleTPCPhaseOut}. However, in mid-2024, Google announced that third-party cookies would remain enabled by default under a user-choice model~\cite{googleTPCPhaseOutDelay}. On 17 October 2025, Google declared that most Privacy Sandbox APIs would be retired in the following months~\cite{privacySandboxPhaseOut}, with deprecation in Chrome M144 (January 2026) and removal in M150 (July 2026)~\cite{privacySandboxDeprecation}.

This sequence of reversals raises a central question: \emph{how was Privacy Sandbox actually used in the Web ecosystem, and how did their adoption evolve during their rise and subsequent downfall?} Prior studies have examined individual APIs---most notably Topics and Protected Audience~\cite{verna2024first, verna2025understanding, beugin2024interest, jha2023robustness, jha2024re, berke2022privacy, thomson2023privacy, alvim2024privacy, philipse2024post, johnson2024unearthing, ali2023navigating, calderonio2024fledging, long2024evaluating, kobayashi2024privacy, thomson2024Protected}. However, no existing study provides a unified, ecosystem-wide, and longitudinal view of Privacy Sandbox usage. Moreover, prior measurements do not account for the role of consent banners and management platforms (CMPs), which critically regulate third-party components execution and may strongly impact when and how these APIs are invoked.

In this paper, we present the first large-scale, longitudinal, and consent-aware measurement of Privacy Sandbox deployment across the Web. Using a custom Chrome-based crawler equipped with a comprehensive API call interceptor, we monitor the usage of the ten Privacy Sandbox APIs that reached general availability. Our methodology explicitly distinguishes between pre-consent, post-consent, and explicit consent-denial execution, allowing us to characterise both intended usage and consent-inconsistent behaviour~\cite{verna2024first, verna2025understanding}.

We conduct weekly crawls of the top-10,000 websites over six months, capturing the evolution of API adoption over time and the third-party ecosystem responsible for their deployment. Our analysis reveals that the contraction of the Privacy Sandbox ecosystem began well before its formal retirement. With the notable exception of partitioned cookies (CHIPS---Cookies Having Independent Partitioned State), adoption remained highly concentrated among a small number of actors and steadily declined throughout our measurement period. Even the APIs that Google plans to maintain exhibit no signs of growth or ecosystem momentum.

Taken together, our findings depict that Google's proposal of a browser-mediated privacy architecture has failed to achieve sustained adoption at Internet scale. {Not only the Privacy Sandbox never took off, but the market started abandoning it well before Google's deprecation announcement.} More importantly, the demise of the Privacy Sandbox leaves unsolved the broader challenge of enabling privacy-preserving, interest-based advertising on the Web and highlights the difficulty of deploying technically complex privacy mechanisms that require substantial changes to entrenched advertising workflows: third-party cookies (and tracking) remain the major technology to fuel the online ad ecosystem {testifying an ecosystem reluctant to abandon stable and working technologies.}.

This paper makes the following contributions.
\begin{itemize}
    \item We provide the first longitudinal, consent-aware measurement of Privacy Sandbox API usage across the Web.
    \item We characterise the adoption dynamics of both retired and maintained APIs, showing that ecosystem disengagement preceded Google’s deprecation decision.
    \item We demonstrate that partitioned cookies (CHIPS) are the only component to achieve widespread deployment, with 50\% of sites using them. Yet, 75\% of websites still use classical persistent unpartitioned third-party cookies.
    \item We document questionable API invocations even when no consent is provided, shedding light on the complex interaction between consent enforcement and actual technical implementation.
\end{itemize}

The rest of the paper is organised as follows: Section~\ref{sec:privacy-sandbox} introduces the Privacy Sandbox initiative and the APIs it includes, while Section~\ref{sec:related-work} provides an overview of related work. Section~\ref{sec:methodology} details the data collection methodology while 
Section~\ref{sec:usageAtLarge} introduces the main results. Next we drill down our investigation: we witness the adoption of CHIPS (or partitioned cookies) in Section~\ref{sec:chips}, while in Section~\ref{sec:deprecated-apis} and Section~\ref{sec:supported-apis} we present the results for retired and maintained APIs, respectively. Section~\ref{sec:questionable} discusses questionable usage of the Privacy Sandbox APIs from the consent point of view. Finally, Section~\ref{sec:conclusion} concludes the paper.

\section{Privacy Sandbox primer}
\label{sec:privacy-sandbox}

The Privacy Sandbox is Google's umbrella initiative to redesign key Web functionalities related to personalised advertising, cross-site tracking, and identity without relying on third-party cookies or other stable cross-site identifiers. Instead of allowing unrestricted access to browser state to third parties, the initiative introduces a set of browser-mediated APIs that constrain data access and relocate privacy-sensitive computation, such as profiling, ad selection, and measurement, into the user’s browser.
A core design principle of the Privacy Sandbox is that the browser acts as the primary enforcement point for privacy. Sensitive operations are executed locally, and only limited, aggregated, or noise-added outputs are released to remote endpoints. In principle, this architecture aims to preserve advertising functionality while reducing the ability of third parties to perform cross-site tracking.

In this work, we focus on the ten APIs that reached general availability in 2023--2024, summarised in Table~\ref{tab:privacy-sandbox} and grouped according to the functional domains defined by Google.\footnote{\label{ps-timeline}\url{https://privacysandbox.com/intl/en_us/open-web/\#the-privacy-sandbox-timeline}, accessed on \today.}

\begin{table*}
    \centering
    \caption{Summary of Privacy Sandbox APIs, grouped by domain. The start quarter refers to the opening period of the first public testing phase.}
    \begin{tabular}{rclccccp{1.7cm}}
        \toprule
        \textbf{API} & & \textbf{Scope} & \textbf{Start} & \textbf{Depr.} & \textbf{Restricted} & \textbf{Ref.} & \textbf{Rel. work}  \\
        \midrule
        \textbf{Attribution Reporting} & & Ad measurement & Q1 '21 & \checkmark & \checkmark & \cite{attributionReportingAPI, attributionReportingAPIExplainer} & \cite{ghazi2025differential} \\
        \midrule
        \textbf{Topics} & \multirow{4}{*}{{\rotatebox[origin=c]{90}{\small \parbox{1.5cm}{\centering Show relevant ads}}}} & Ad personalisation & Q2 '22 & \checkmark & \checkmark & \cite{topicsAPI, topicsAPIExplainer} & \cite{verna2024first, verna2025understanding, beugin2024interest, jha2023robustness, jha2024re, berke2022privacy, thomson2023privacy, alvim2024privacy} \\
        \textbf{Protected Audience} & & Ad retargeting & Q2 '22 & \checkmark & \checkmark & \cite{prAuAPI, prAuAPIExplainer} & \cite{philipse2024post, johnson2024unearthing, ali2023navigating, calderonio2024fledging, long2024evaluating, kobayashi2024privacy, thomson2024Protected} \\

        \midrule
        \textbf{Private Aggregation} & 
        \multirow{6}{*}{{\rotatebox[origin=c]{90}{\small \parbox{2.5cm}{\centering Cross-site privacy boundaries}}}} 
        & Privacy-preserving data aggregation & Q1 '22 & \checkmark & \checkmark & \cite{privateAggregationAPI, privateAggregationAPIExplainer} & \cite{ghazi2025differential} \\
        \textbf{Shared Storage} & & Cross-site data storage and aggregation &  Q2 '22 & \checkmark & \checkmark & \cite{sharedStorageAPI, sharedStorageAPIExplainer} & \cite{nisenoff2025exploiting} \\
        \textbf{Related Website Sets} &
        & Cross-site identity and storage grouping & Q2 '21 & \checkmark & - & \cite{rwsAPI, rwsAPIExplainer} & \cite{mcquistin2024first} \\
        \textbf{CHIPS} & & Opt into partitioned cookies & Q1 '22 & - & - & \cite{chipsAPI, chipsAPIExplainer} & \cite{zollner2025first} \\
        \textbf{Fenced Frames} & & Privacy-preserving Iframe rendering & Q2 '22 & - & - & \cite{fencedFramesAPI, fencedFramesAPIExplainer} & - \\
        \textbf{FedCM} & & Identity and single sign-on & Q2 '22 & - & - & \cite{fedCMAPI, fedCMAPIExplainer} & - \\

        \midrule
        \textbf{Private State Tokens} & & Spam and fraud prevention & Q1 '21 & - & - & \cite{pstAPI, pstAPIExplainer} & \cite{ali2023navigating} \\
        \bottomrule
    \end{tabular}
    \label{tab:privacy-sandbox}
\end{table*}

\subsection{Advertising and measurement APIs}

\paragraph{Attribution Reporting API \cite{attributionReportingAPI, attributionReportingAPIExplainer}}
It enables privacy-preserving measurement of \textit{ad conversions} by linking user interactions (e.g., ad clicks or views) with subsequent actions on an advertiser's site without exposing cross-site identifiers. The browser generates event-level or aggregated reports locally and releases only limited, noise-added signals to reporting endpoints.

\paragraph{Topics API \cite{topicsAPI, topicsAPIExplainer}}
Provides a privacy-preserving mechanism for \textit{interest-based advertising}, allowing advertisers to obtain the interests of a user without learning about the specific websites visited. It represents the evolution of the discontinued FLoC proposal \cite{floc}. The browser locally derives coarse-grained \textit{topic identifiers} based on recent browsing history. During ad selection, the browser discloses some of the topics to the advertiser, occasionally introducing random noise to limit traceability.

\paragraph{Protected Audience API \cite{prAuAPI, prAuAPIExplainer}}
Formerly \textit{FLEDGE}, it enables privacy-preserving ad \textit{re-targeting}, i.e., the ability of displaying ads about websites previously visited. When a user visits a website, advertisers can assign them to an \emph{interest group}, representing the site or a product. During ad rendering, the browser runs an auction among eligible interest groups and fetches the winning ad. Auction reports and related metadata are generated by the browser and released under controlled channels.

\paragraph{Private Aggregation API \cite{privateAggregationAPI, privateAggregationAPIExplainer}}
It enables the generation of aggregated statistics based on client-side data, without revealing individual user information. It is designed to be used in combination with other APIs (Protected Audience and Shared Storage) to compute metrics such as ad reach and auction winning rate while preserving user anonymity.

\subsection{Storage, identity, and partitioning APIs}

\paragraph{Shared Storage API \cite{sharedStorageAPI, sharedStorageAPIExplainer}}
This API introduces a new form of shared browser storage that can be accessed across multiple domains. To prevent data leakage, the stored data can only be accessed from a secure JavaScript environment known as a \emph{worklet}. This design enables privacy-preserving use cases, such as frequency capping or A/B testing, without exposing raw user-level data to external scripts.

\paragraph{Related Website Sets API \cite{rwsAPI, rwsAPIExplainer}}
Formerly \textit{First-Party Sets}, it provides a way for organisations to declare multiple domains as part of a single group of services. Browsers apply looser privacy boundaries within a set—for example, enabling shared cookies or identity workflows, while still isolating unrelated third parties. The mechanism targets large multi-domain services. (e.g., \url{shop.acme.com} and \url{support.acme.co.uk}).

\paragraph{CHIPS (Cookies Having Independent Partitioned State) \cite{chipsAPI, chipsAPIExplainer}}
Provides a standardised way to mark cookies as \textit{partitioned} by the first-party site. Each third-party embedded in a website receives an isolated cookie jar, preventing cross-site reuse of identifiers while preserving legitimate embedded functionality (e.g., login widgets, shopping carts, media players).

\paragraph{Fenced Frames API \cite{fencedFramesAPI, fencedFramesAPIExplainer}}
Introduces a \texttt{<fencedframe>} HTML element, which embeds content similarly to an \texttt{iframe} but prevents data sharing between the top-level site and the embedded content (e.g., an ad banner). This allows ads or other third-party content to be rendered securely and privately, without exposing user data or permitting bidirectional communication.

\subsection{Identity and security APIs}
\paragraph{Federated Credential Management API (FedCM) \cite{fedCMAPI, fedCMAPIExplainer}}
Allows users to log into sites using their federated identity (e.g., ``Sign in with...'') without relying on third-party cookies or navigational redirects, normally used by protocols such as OAuth 2.0. By mediating authentication exchanges directly within the browser, FedCM reduces cross-site tracking risks while preserving usability for both users and identity providers.

\paragraph{Private State Tokens API \cite{pstAPI, pstAPIExplainer}}
Formerly \textit{Trust Tokens}, it allows websites to issue cryptographically signed tokens that \textit{attest to a user's legitimacy} (e.g., ``human vs. bot'' CAPTCHA) without exposing identity or linking user's activity across sites. This minimises the need for passive user tracking, such as that used by other CAPTCHA services.

\subsection{Deployment characteristics}

A distinctive feature of the Privacy Sandbox ecosystem is that several APIs require explicit developer enrolment before use (marked as ``Restricted'' in Table~\ref{tab:privacy-sandbox}. Third parties must complete an attestation process with Google to obtain the necessary permissions, which the browser verifies at runtime. This mechanism limits API access to a vetted subset of actors and directly affects observable adoption in the wild. These restricted APIs are related to the processing of privacy-sensitive data.

As of early 2026, 265 organisations have completed the enrolment process for some restricted API, up from approximately 220 reported in March 2025. However, enrolment does not imply sustained or production-level deployment. In the remainder of this paper, we show that actual usage remains highly concentrated among a small subset of these actors and that many enrolled entities never invoke the APIs at scale.

Following Google's October 2025 announcement~\cite{privacySandboxPhaseOut}, only four of the ten APIs described above---CHIPS, FedCM, Private State Tokens, and Fenced Frames---are expected to remain supported, while the others are scheduled for removal in July 2026.

\section{Related work}
\label{sec:related-work}

Research on Google’s Privacy Sandbox spans three partially overlapping directions: 
(i) analyses of the privacy and security properties of individual APIs, 
(ii) empirical measurements of their deployment and adoption in the wild, and 
(iii) studies of adjacent or competing privacy-preserving Web technologies.
We review these lines of work below and position our contribution accordingly, with Table \ref{tab:privacy-sandbox} that summarises the main works grouped by API.

\paragraph{Privacy and security analyses of individual APIs.}
The majority of prior work has focused on the \emph{Topics API} and its predecessor FLoC. Several studies have shown that, despite coarse-grained categorisation and built-in randomisation, Topics disclosures can be aggregated over time to probabilistically re-identify users or link their activity across sites~\cite{jha2023robustness, jha2024re, beugin2024interest, thomson2023privacy, alvim2024privacy}.
Earlier work on FLoC similarly demonstrated that cohort identifiers could be exploited for cross-site inference~\cite{berke2022privacy}.
Together, these results highlight the intrinsic tension between interest-based advertising and resistance to cross-site tracking, even in the absence of explicit identifiers.

The \emph{Protected Audience API} (formerly FLEDGE) has received comparable scrutiny. Multiple works demonstrate that covert channels and protocol-level artefacts can be exploited to leak user information
or construct stable identifiers, undermining the intended privacy guarantees~\cite{long2024evaluating, ali2023navigating, calderonio2024fledging, thomson2024Protected}.
These findings question whether browser-mediated ad auctions meaningfully limit tracking in adversarial settings.

Other Privacy Sandbox components have attracted more limited attention.
Ghazi \emph{et al.} formally analyse the differential privacy guarantees of the
\emph{Attribution Reporting} and \emph{Private Aggregation APIs}, showing that their reporting mechanisms satisfy
differential privacy even under interactive use~\cite{ghazi2025differential}.
In contrast, Nisenoff \emph{et al.} demonstrate that the \emph{Shared Storage API} can be abused as a covert channel,
enabling cross-site information leakage despite its restricted execution model \cite{nisenoff2025exploiting}.
McQuistin \emph{et al.} examine the assumptions underlying \emph{Related Website Sets}, showing that users frequently
fail to correctly identify site affiliation, potentially enabling overly permissive cross-domain data sharing \cite{mcquistin2024first}.
Finally, early implementation-focused work on \emph{CHIPS} reported limited adoption among tracking domains during
initial deployment phases \cite{zollner2025first}.

\paragraph{Measurement studies of Privacy Sandbox deployment.}
Several empirical studies have measured the adoption of individual Privacy Sandbox APIs during early deployment. Large-scale crawls report that the Topics API appeared on approximately 30-40\% of websites during testing phases,
but with fragmented and inconsistent usage patterns, often indicative of experimentation or A/B testing~\cite{johnson2024unearthing,verna2024first,verna2025understanding}.
These studies also highlight frequent invocation before user consent or through unauthorised scripts, commonly attributed to misconfigured consent management logic~\cite{verna2025understanding}.

Measurements of the Protected Audience API consistently report much lower adoption. Early work observed usage on roughly 1\% of websites \cite{philipse2024post}, with later measurements showing peak penetration around 25\% during intensive testing periods~\cite{johnson2024unearthing}.
Importantly, adoption was found to be heavily concentrated on Google-controlled infrastructure, with the vast majority of auctions initiated by Google-owned domains~\cite{calderonio2024fledging}.

To date, no deployment-focused studies provide a unified view across \emph{all} Privacy Sandbox APIs, nor do existing measurements analyse adoption trends over extended time periods or explicitly model the role of consent management platforms in regulating API invocation. {Among the present and past attempts, a now-offline\footnote{\url{https://web.archive.org/web/20240602062407/https://pscs.glitch.me/}, accessed on \today} system used to monitor the usage of some Privacy Sandbox API, only up to September 2024 when the system was abandoned. Conversely, Google Chrome's own popularity dashboard \footnote{\url{https://chromestatus.com/metrics/feature/popularity}, accessed on \today} offers a too coarse level of aggregation, making direct comparison difficult. In both cases, our results appear in line with those shown by such tools.}

\paragraph{Positioning of this work.}
Our work differs from prior literature in three fundamental respects. First, rather than analysing individual APIs in isolation, we provide a \emph{comprehensive, cross-API} measurement of all ten Privacy Sandbox components that reached general availability.
Second, we conduct a \emph{longitudinal analysis} spanning more than six months, capturing both the expansion and contraction phases of deployment preceding Google’s deprecation announcement.
Third, our measurements are explicitly \emph{consent-aware}, distinguishing pre-consent, post-consent, and post-denial execution contexts, which prior studies largely overlook despite their central role in regulating third-party behaviour.

By combining these dimensions, our work complements existing privacy and security analyses with ecosystem-level evidence about deployability, concentration of adoption, and the practical limits of browser-enforced privacy architectures in the contemporary Web advertising ecosystem.

\section{Data collection and datasets}
\label{sec:methodology}

\newcommand{\beforeA}{\textit{Before-Accept}\xspace}
\newcommand{\beforeD}{\textit{Before-Deny}\xspace}
\newcommand{\beforeC}{\textit{Before-Click}\xspace}
\newcommand{\afterA}{\textit{After-Accept}\xspace}
\newcommand{\afterD}{\textit{After-Deny}\xspace}

To characterise the deployment of Privacy Sandbox APIs in the wild, we rely on large-scale, longitudinal active crawling. We design and implement a custom measurement framework that instruments Google Chrome to detect invocations of Privacy Sandbox APIs at runtime, while explicitly modelling user consent states that govern third-party execution on modern websites.

Our measurement campaign targets the top-10,000 websites according to the Tranco ranking and runs weekly over a six-month period. For each site and each crawl iteration, we emulate realistic consent interactions and distinguish between execution occurring before consent, after explicit consent, and after explicit consent denial. This design allows us to capture both intended post-consent usage and execution that is inconsistent with the user’s declared choice.

At a technical level, our crawler combines browser-level instrumentation via the Chrome DevTools Protocol with JavaScript-level interception to observe API invocations across the main document and embedded frames. In addition to Privacy Sandbox calls, we record third-party requests and cookie-setting behaviour, enabling joint analysis of API adoption and cookie partitioning.

\subsection{API interceptor}
\label{sec:call-listener}
The goal of our instrumentation is to detect runtime invocations of Privacy Sandbox APIs as they occur in the browser, together with their execution context and caller origin, while preserving the original semantics of page execution. We aim to capture API usage across the main document and embedded frames, independently of whether calls originate from first-party or third-party scripts.

We build on and extend Priv-Accept~\cite{verna2025understanding} to record the Privacy Sandbox API usage using two complementary approaches. In both cases, we use the Selenium library to communicate with the browser through the Chrome DevTools Protocol (CDP)~\cite{googleCDP}, which allows an external program to issue fine-grained commands and listen to specific events.
When available, we rely on native Chrome DevTools Protocol (CDP) events, as they provide direct, browser-level visibility into API activity and are not affected by JavaScript obfuscation or redefinition.
With this method, we can track 3 Privacy Sandbox APIs: Attribution Reporting, FedCM and Fenced Frames (e.g., \texttt{Storage.attributionReportingSourceRegistered}\footnote{\url{https://chromedevtools.github.io/devtools-protocol/tot/Storage/\#event-attributionReportingSourceRegistered}, accessed on \today}) and record all the exposed parameters. 

For APIs that do not emit CDP events, we instrument their JavaScript entry points to observe invocation at runtime, a technique commonly used in Web measurement to capture dynamic API usage.
To this end, we use the CDP method \texttt{Page.addScriptToEvaluateOnNewDocument} to inject a custom script into the page~\cite{senol2023unveiling}. The script automatically wraps selected functions (e.g., \texttt{document.browsingTopics()} for the Topics API). Each wrapper logs the function name, arguments, return value, and script origin, and forwards this information to the interceptor using CDP bindings (\texttt{Runtime.addBinding}).\footnote{\url{https://chromedevtools.github.io/devtools-protocol/tot/Runtime/\#method-addBinding}, accessed on \today.} The original API is finally executed to avoid altering the behaviour of the webpage.
We inject the script into the main page and every iframe as soon as they are discovered.
Iframe discovery is facilitated by CDP's ``auto-attach'' feature,\footnote{\url{https://chromedevtools.github.io/devtools-protocol/tot/Target/\#method-setAutoAttach}, accessed on \today.} with the \texttt{waitForDebuggerOnStart} option enabled. This option forces the browser to halt all activity (i.e., page loading and script execution) whenever a new iframe is found and transfer the control to the CDP until it completes its task.\footnote{\url{https://chromedevtools.github.io/devtools-protocol/tot/Runtime/\#method-runIfWaitingForDebugger}, accessed on \today.}
This allows us to proceed with the script injection before any iframe element loads, ensuring that we do not miss any invocations due to race conditions.
All wrappers transparently forward arguments to the original API implementation and return its original return value, ensuring that page behaviour is not altered by our instrumentation.

Thanks to this twofold strategy, we can collect, for each API of the Privacy Sandbox, the name of the function being called, the URL of the caller, henceforth \emph{Calling Party} (CP)\footnote{We define the calling party (CP) as the eTLD+1 of the script origin responsible for invoking the API; for inline scripts, we associate the call with the document origin.}, the domain of the frame where the call occurred, eventual parameters passed to the function and its return value, independently of the context where the call happens.

\subsection{Managing consent banners: \beforeA, \afterA, \afterD visits}


Since Privacy Sandbox APIs are predominantly invoked by third-party scripts whose execution is often gated by consent management platforms (CMPs), any measurement that ignores the user's consent risks substantially mischaracterising both the prevalence and timing of API usage~\cite{jha2022internet,verna2025understanding}. For this reason, our crawler models three distinct user-consent states and performs separate visits in each state to capture the Privacy Sandbox behaviour under realistic CMP-controlled loading logic.

\textbf{\beforeA.}
For every website, the crawler performs an initial visit with a fresh browser profile. The crawler does not interact with the page and only records all Privacy Sandbox API invocations. During this phase, no consent is provided: any API invocation here reflects code executed before user agreement, either because CMPs fail to block components correctly or because the website does not implement blocking at all. The Before-Accept state approximates the experience of a first-time user who has not yet interacted with the consent banner, and API invocations reflect code executed before any explicit user choice.

\textbf{\afterA}
After completing the first visit, we attempt to locate and click an ``Accept button'' by matching it against a curated list of multilingual keywords.\footnote{We include keywords in English, French, German, Italian and Spanish, following the approach of Priv-Accept~\cite{jha2022internet}.} After accepting consent, we clear the browser cache and reload the page to ensure that all consent-dependent scripts are executed under the new state, yielding a clean and comparable post-consent measurement. This yields a second measurement where all components released after consent, including additional ads, profiling scripts, and consent-dependent Privacy Sandbox calls, can execute. If no Accept button is found, we skip the \afterA visit. We treat the After-Accept state as the reference scenario for legitimate Privacy Sandbox deployment, when the users granted consent to API usage.

\textbf{\beforeD and \afterD}.
To complement the process, we extend the crawler to emulate users explicitly denying consent. Replicating the same strategy, we collect a list of keywords to search for a ``Deny button'' within a privacy banner. We manually build the list by visiting the top-100 websites according to the SimilarWeb~\cite{similarweb} list in Italy, Germany, France, Spain and the UK. The final list contains 166 keywords across the five languages.

During this analysis, we verified that many banners do not expose a straightforward ``Deny Button'', despite regulatory guidance (e.g., 
France's data protection authority CNIL, which mandates that rejecting cookies must be as simple as accepting them): some hide the control behind ``Options'' submenus, while others present a ``Deny and Subscribe'' paywalled alternative. Unlike simpler crawlers, our implementation supports multi-step denial flows in which the “deny” option is nested behind configuration or “options” menus, increasing coverage of realistic CMP designs. However, it cannot bypass paywalls.

The resulting \afterD dataset captures Privacy Sandbox activity when the user has explicitly refused personal-data processing. Any invocation observed here is labelled as \emph{questionable} behaviour, in line with~\cite{verna2025understanding}, regardless of their legal interpretation.

In the remainder of the paper, we collectively refer to \beforeA and \beforeD as \beforeC, and to any API invocation observed in \afterD as consent-inconsistent usage. In summary, we have:
\begin{itemize}
    \item Pre-consent execution (\beforeA/\beforeD, collectively \beforeC);
    \item Legitimate consent-dependent usage (\afterA);
    \item Consent-inconsistent usage (\afterD).
\end{itemize}

\subsection{Data collection}
\label{sec:data-collection}

To provide a complete picture of Privacy Sandbox usage over time, we run the crawler to visit the top-10,000 websites according to the Tranco list~\cite{lepochat2019tranco} that we refresh weekly to reflect current site popularity.\footnote{Tranco provides only a single global ranking, but it is the most established ranking service, while SimilarWeb offers a per-nation list but only for the top-100 websites.}
All crawls are executed using Google Chrome version 138.0.7204.4, with all Privacy Sandbox features explicitly enabled. We fix the browser version throughout the measurement period to avoid confounding effects due to browser updates.

For each website, we store:
\begin{itemize}
    \item the URL of each first- and third-party object downloaded by the browser to render the website;
    \item the set of cookies created during the visit;
    \item the calls made to each Privacy Sandbox API.
\end{itemize}
For each cookie, we collect its name, value and other relevant metadata, such as the domain of its creator and whether it is partitioned. 

We repeat the crawl weekly, starting every Monday at 00:00 (GMT+1) from a European server. For each website, we load the landing page (following redirects when present) up to four times to capture the \textit{before} and \textit{after} consent states, allowing up to 60\,s per visit. To mitigate transient network failures and rate limiting, failed visits are retried up to three times before being discarded for that snapshot.
The campaign runs from 23 June 2025 to 5 January 2026, covering more than six months. Using 20 crawler instances in parallel, each full cycle completes in roughly 36 hours.

As with any automated crawl, websites or embedded third parties may detect non-human behaviour and alter page execution, for example, by presenting CAPTCHA or verification interstitials or by suppressing third-party script execution (cloaking). Likewise, the crawler may occasionally fail to trigger the ``Accept/Deny’’ flow correctly. These effects introduce randomness and potential bias into our measurements.

We mitigate their impact by repeating measurements weekly and analysing trends across large populations rather than individual websites. Furthermore, we compute all statistics over the set of websites successfully visited for the corresponding visit type and week. This mitigates bias due to population changes, such as the fact that the \afterA and \afterD datasets are smaller than the \beforeC dataset.

Finally, our primary findings rely on persistent adoption patterns and strong concentration effects among major third parties, which are inherently less sensitive to sporadic cloaking or transient crawl failures.

\subsection{Dataset overview}
\label{sec:dataset}

\begin{table}[]
\caption{Summary of the number of successful visits for the different paths followed by the crawler.}
\begin{tabular}{cccr}
\toprule
\textbf{Visit}          & \textbf{Action} & \multicolumn{1}{l}
{\textbf{Visits min-max}} & \multicolumn{1}{c}{\textbf{3\textsuperscript{rd} parties}} \\
\midrule
\multirow{3}{*}{\rotatebox[origin=c]{90}{\textit{Before}}} & \textit{Accept} & 7,625--8,262 & 7,324--7,973 \\
                        & \textit{Deny}   & 7,423--8,448 & 7,041--7,989 \\
                        \cmidrule(lr){2-4}
                        & \textit{Click}  & 7,703--8,473 & 7,354--8,025\\ \midrule
\multirow{2}{*}{\rotatebox[origin=c]{90}{\textit{After}}}  & Accept & 3,209--3,772 & 4,471--5,132 \\
                        & \textit{Deny}   & 2,082--2,199 & 2,524--2,906 \\ \bottomrule
\end{tabular}
\label{tab:crawler sum up}
\end{table}

We present a summary of first and third parties contacted in the different scenarios of the crawling activity in Table~\ref{tab:crawler sum up}. 
For completeness, we report the \beforeA and \beforeD visits separately, and their merging in the \beforeC dataset. Here, we also merge the set of contacted third parties, which is the union of the contacted third party sets in \beforeA and \beforeD.

During each week, we successfully visited between 7,400 and 8,400 websites in any \textit{Before-} visit. On average, 93\% of websites overlap between the \beforeA and \beforeD visits across all dates. 
The high overlap between \beforeA and \beforeD visits indicates that failures are largely stochastic and do not systematically affect specific subsets of websites, e.g., random connection failures and website anti-bot measures like CAPTCHA.
Interestingly, in the \beforeC visit, each website includes fewer than one third-party request on average.

Moving to the \afterA visits, our crawler succeeds in accepting the website policy in about 44\% of \beforeA visits, a percentage in line with \cite{jha2022internet}. The majority of failures are due to the website not showing a consent banner at all. During the \afterA visits, we encounter about 4,400-5,100 total third parties, with the average number of third parties per website that grows to 1.4-1.5. This reflects the fact that websites enable additional third-party-powered features, e.g., ads, after users provide their consent.

For the \afterD, the crawler succeeds in denying consent in about 2,100 cases ($\approx$28\% of the successful \beforeD cases). In addition to the cases where no banner is present, this lower percentage is due to the additional complexity of denying cookies, which often requires multiple interactions or access via paywalled mechanisms. 

To better characterise the limitations of automated denial, we manually inspected a random sample of 50 websites where the crawler failed to deny consent. In 23 of them, there is no banner, 7 are in unsupported languages, 6 use multiple toggles or checkboxes instead of a single deny button, 5 are not active, 2 have no deny option, and 1 offers a paywalled option. In the remaining cases, the crawler failed to match the keywords (3 sites), or was unable to detect the deny button despite its presence (3 sites).
In short, failures are primarily due to the absence of a consent banner, unsupported languages, multi-toggle interfaces, hidden deny options, or paywalled alternatives.

\begin{figure}
    \centering
    \includegraphics[width=\linewidth]{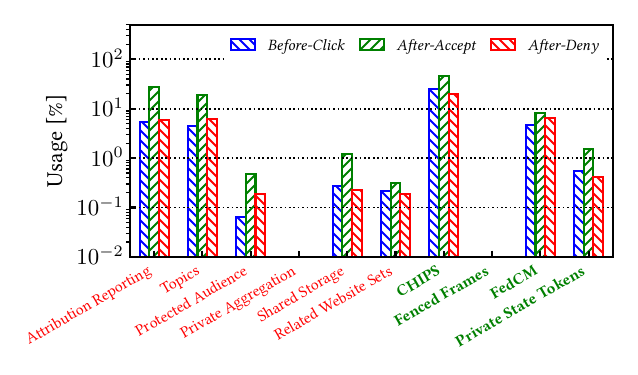}
    \caption{Percentage of websites where each API was called during the 5 January 2026 snapshot. In red, the retired APIs. In bold green, the still-supported ones.}
    \label{fig:usage-per-api-visit}
\end{figure}

In the following sections, we leverage these datasets to characterise Privacy Sandbox usage at scale, first providing a cross-API snapshot and then analysing longitudinal adoption trends.
To support reproducibility, the code and datasets used in this study are released as described in Appendix~\ref{app:ethics}.

\section{Privacy Sandbox usage at large}
\label{sec:usageAtLarge}
We begin by providing a cross-API snapshot of Privacy Sandbox usage using the most recent measurement week, in order to characterise overall adoption levels and the impact of consent state on observed API invocation.
Figure~\ref{fig:usage-per-api-visit} reports, for each API, the share of websites where at least one invocation is recorded, distinguished by consent state. Recall that we compute percentages over the set of successfully visited websites for that specific visit type to avoid bias.

Figure~\ref{fig:usage-per-api-visit} reveals a clear stratification among Privacy Sandbox components, with a small number of APIs exhibiting non-negligible deployment and the majority remaining marginal.
Only CHIPS, Topics, and Attribution Reporting API exceed the 10\% threshold for \afterA visits. CHIPS (active on $\approx 50$\% of websites)  stands apart from other components, as it enables stateful embedded functionality rather than user profiling, and its usage is therefore not necessarily contingent on user consent.
For several restricted APIs, we observe invocations in \beforeC and, more notably, \afterD visits, indicating execution that is inconsistent with the declared user choice. As shown in prior work, such behaviour often stems from CMP misconfiguration or third-party integration errors rather than deliberate misuse~\cite{verna2025understanding}. We will investigate this later.

\subsection{\afterA usage}
We focus on the \afterA state as it represents the intended deployment scenario for Privacy Sandbox APIs, where execution is explicitly enabled by user consent. Here, we observe the uneven adoption landscape (green bars in Figure~\ref{fig:usage-per-api-visit}). After CHIPS, Topics API ($\approx$17\%) and Attribution Reporting API ($\approx$21\%) show notable penetration. All remaining components show marginal use: FedCM reaches barely 6\%, while Protected Audience, Private State Tokens, and Related Website Sets usage remain below 1\%.\footnote{The Protected Audience figure is in line with Google's statement in~\cite{privacySandboxDeprecation}.} Private Aggregation and Fenced Frames are practically absent. In the following, where not explicitly indicated, the figures show API penetration in the \afterA.

Overall, these figures indicate limited ecosystem uptake of most Privacy Sandbox APIs, even two years after they reached general availability.
Given its markedly higher prevalence, we analyse CHIPS in detail in Sec.~\ref{sec:chips} and compare its deployment to that of traditional third-party cookies.

\subsection{Effect of API retirement}
\begin{figure}[t]
    \centering
    \includegraphics[width=\linewidth]{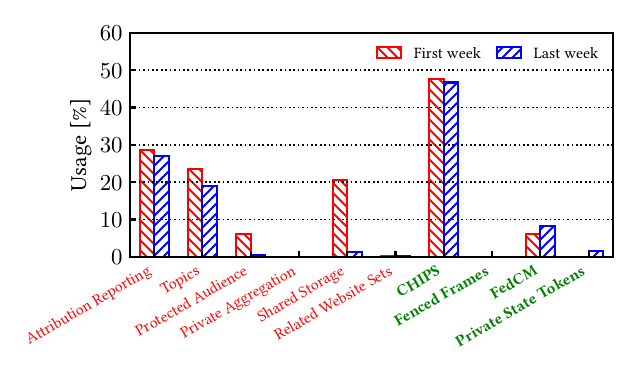}
    \caption{Percentage of websites where each API call was encountered in the first and last week of measurements, \afterA. In red, the retired APIs. In bold green, the still-supported ones.}
    \label{fig:usage-first-last-week}
\end{figure}

Figure \ref{fig:usage-first-last-week} compares API presence during the first and last measurement week (23 June $\to$ 5 January) in the \afterA. The overarching trend is a contraction:
for retired APIs, Shared Storage and Protected Audience collapse almost entirely, falling from $\approx$20\% and $\approx$8\% to near zero. Topics and Attribution Reporting decline more gradually (-1.5 to -4.5 percentage points), and their usage remains non-negligible at 19–27\%.

Among maintained APIs (in green), CHIPS remains the most pervasive, maintaining a stable adoption rate slightly below 50\% of the websites. FedCM and Private State Tokens remain low and flat across the entire period. Fenced Frames are ignored altogether.

Crucially, Google’s announcement that these APIs would be maintained did not coincide with any visible increase in their adoption, suggesting a broader retreat from Privacy Sandbox technologies rather than a reallocation of adoption efforts.

\begin{figure}[t]
    \centering
    \includegraphics[width=\linewidth]{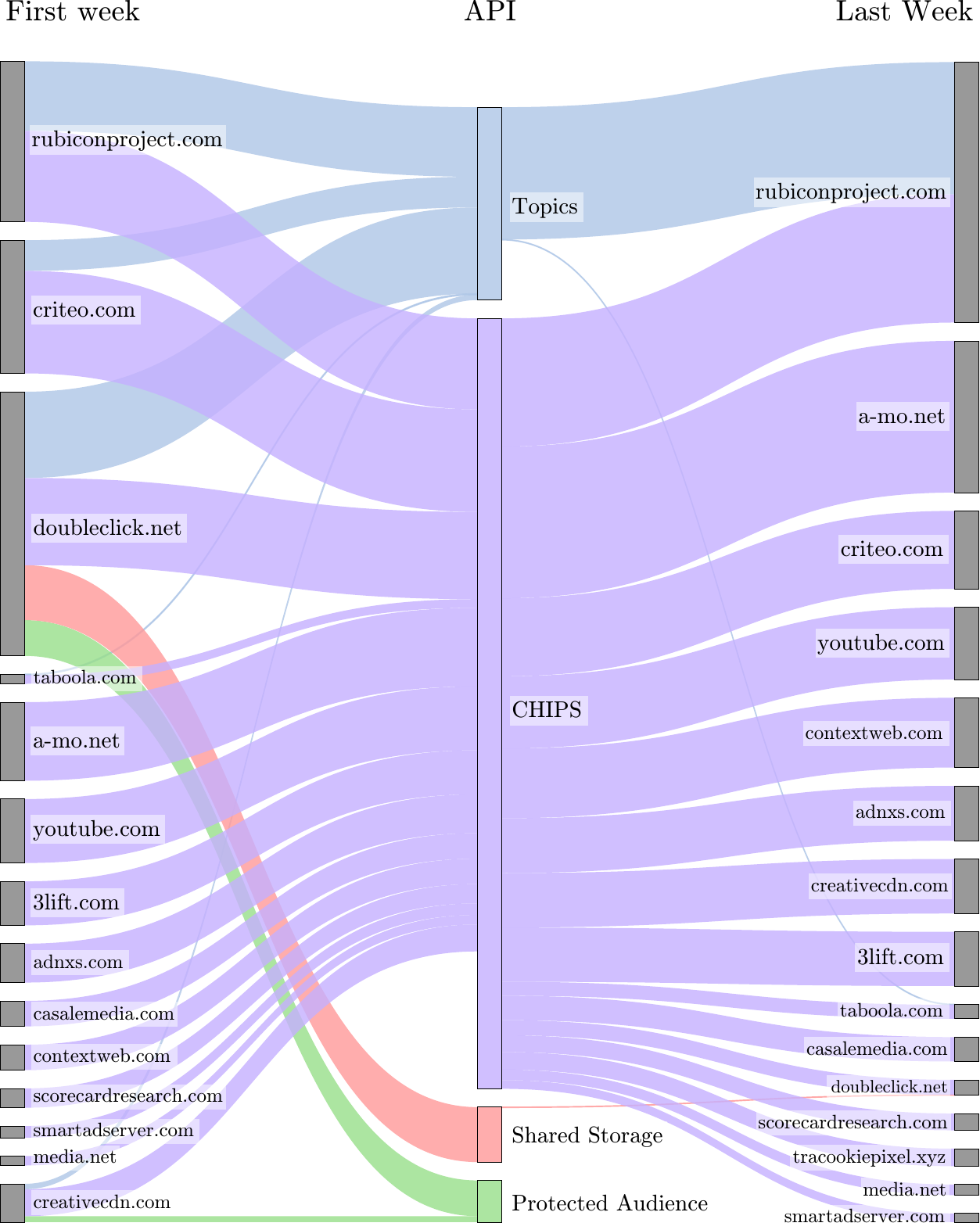}
    \caption{Flow graph of third-party callers during the latest campaign.}
    \label{fig:third-party-callers}
\end{figure}

\subsection{Adopter Third Parties}

Out of more than 7,000 third parties observed in our data, 265 are attested entities. Only 50 of them called at least one Privacy Sandbox API during the first week; callers drop to 41 by January 2026 (see Section~\ref{sec:deprecated-apis} and \ref{sec:supported-apis} for details), testifying how the set of adopters is extremely concentrated. 

Figure~\ref{fig:third-party-callers} shows the top 15 calling domains among the 41 still active in January. We compare their API usage with that seen during the first week of our measurements. The thickness of the link between the CP and the API represents the number of websites where a third party calls an API. The landscape is dominated by major advertising intermediaries such as \url{doubleclick.net}, \url{criteo.com}, and \url{rubiconproject.com}. All rely on CHIPS, often exclusively. Topics API remains their second most common choice, although only \url{rubiconproject.com} uses it consistently. It is interesting to note that, although the overall CHIPS adoption rate remains stable as reported in Figure~\ref{fig:usage-first-last-week}, the usage among the most pervasive third parties increases.

Shared Storage and Protected Audience disappear rapidly: \url{doubleclick.com}---the last relevant user of Shared Storage---withdraws it towards the end of the measurement period, causing the API’s usage to collapse to almost 0 across all websites.

Interestingly, Attribution Reporting is under-represented in this analysis because many events expose only the execution context and not the initiating third party, limiting caller attribution. This is an artefact of the API’s reporting model that does not allow us to trace back those invocations to the triggering third-party platform.

Curiously, \url{doublelick.net}---the Google-owned digital advertising platform---was the top adopter in the first week. In the last week, it shows the greatest usage reduction, abandoning all APIs but CHIPS. \url{creativecdn.com}, the only other adopter of the Protected Audience, abandons them altogether. Conversely, \url{criteo.com} and \url{rubiconproject.com} maintain significant usage of the Topics API.
{This trend is confirmed in Figure~\ref{fig:third-party-callers-alt}, which gives a caller-centric view of API usage.}

Taken together, these results depict an ecosystem in which only a small subset of Privacy Sandbox components achieved measurable deployment, and even those exhibit declining or stagnant adoption. In the remainder of the paper, we examine these trends in greater detail, first analysing CHIPS and cookie partitioning (Sec.~\ref{sec:chips}), and then focusing on the longitudinal evolution of retired and maintained APIs (Sec.~\ref{sec:deprecated-apis} and \ref{sec:supported-apis}).

\section{{Cookies Having Independent Partitioned State (CHIPS)}}
\label{sec:chips}

In this section, we focus on the adoption of CHIPS, the most widely deployed Privacy Sandbox component, and we compare it to traditional, unpartitioned third-party cookies. Although Google presents CHIPS as a Privacy Sandbox API, it reduces to setting the \texttt{Partitioned} attribute in the HTTP Set-Cookie header (or in the equivalent JavaScript API). Cookies marked as \texttt{Partitioned} are stored in an isolated environment per top-level site, so that a third party cannot re-access the same cookie value across different first parties, thereby preventing cross-site tracking and the possibility of building browsing profiles. In the remainder of this paper, we refer to cookies with the Partitioned attribute as partitioned cookies, and to all others as traditional cookies.

\subsection{CHIPS v. traditional cookies}
\label{sec:chips-v-trad-cookies-accept}

\begin{figure}
    \centering
    \includegraphics[width=\linewidth]{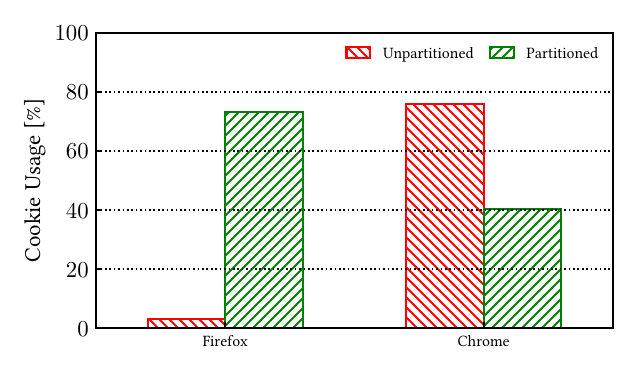}
    \caption{Percentage of websites where a third party installs a cookie (both session and persistent cookies).}
    \label{fig:cookie-usage-after-accept}
\end{figure}

We first dig deeper into the comparison between partitioned and traditional cookies, comparing the rate of websites where third parties install cookies. To make a greater sense of this comparison, we also consider a similar crawling procedure on Mozilla Firefox, which enforces cookie partitioning by default since 2021.\footnote{\url{https://blog.mozilla.org/security/2021/02/23/total-cookie-protection/}, accessed on \today.} {We include Firefox since the browser only supports the CHIPS API: this offers us the possibility to compare two different approaches: a CHIPS-first (Firefox) and a CHIPS-optional one (Chrome).} We apply the same methodology described in Section~\ref{sec:methodology}, including the Firefox browser (ver. 146.0.1) and executing exactly the same accept-or-deny-cookies stages. In total, with Firefox, we were able to click on an ``Accept''-or-equivalent button in 3,409 websites, and a ``Deny''-or-equivalent in 2,239 websites, figures that are comparable with the Google Chrome scenario. For this analysis, we consider both session and persistent cookies.

Figure~\ref{fig:cookie-usage-after-accept} sums up the comparison, showing clear differences: using Chrome, we observe a third party installing a partitioned cookie in around 40\% of websites, while in 70\% of cases they still install unpartitioned cookies. With Firefox, third parties are forced to use partitioned cookies only, reducing the business opportunities of tracking. Note indeed that the percentage of websites where we observe a traditional cookie in Chrome is the same as where we observe partitioned cookies in Firefox.

Finally, some unpartitioned cookies are observed in Firefox. A manual inspection reveals that these cases arise from limitations in our cookie classification: we classify cookies as third-party when their effective TLD differs from that of the visited site. In some cases, TLDs slightly differ (e.g., cookies with domain \url{zoom.com} set on \url{zoom.us}), even though the two domains belong to the same service.\footnote{We use the \textit{tldextract} library \url{https://github.com/joeguo/tldextract} and the DuckDuckGo map \url{https://github.com/duckduckgo/tracker-radar/blob/main/build-data/generated/domain_map.json} to extract the TLD.}
Correctly identifying these cases would require manual and domain-to-service reconciliation. We therefore acknowledge the presence of this bias in our dataset ($\approx$2.5\% of Firefox websites contain seemingly third-party cookies), which may similarly affect Google Chrome.

\subsection{CHIPS v. other Privacy Sandbox APIs}

\begin{figure}
    \centering
    \includegraphics[width=\linewidth]{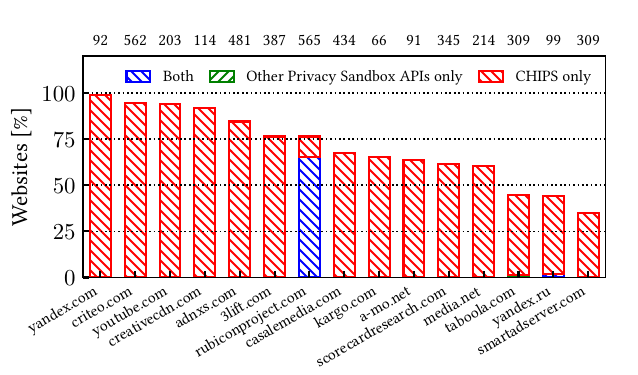}
    \caption{Percentage of websites where a third party calls a Privacy Sandbox API other than CHIPS, in \afterA. The top axes contains the amount of websites where each third party is active.}
    \label{fig:chips-vs-others-accept}
\end{figure}

We then move to comparing the usage of partitioned cookies with that of the other Privacy Sandbox APIs. We do so in Figure~\ref{fig:chips-vs-others-accept}, listing the 15 third parties using a Privacy Sandbox technology in the largest share of websites---among those present on at least 50 websites. 
Among highly prevalent third parties, CHIPS is the only Privacy Sandbox technology in use. Only \url{rubiconproject.com} uses other Privacy Sandbox APIs (Topics API in particular, as shown in Figure~\ref{fig:third-party-callers}, combining them with CHIPS. \url{yandex.com} installs partitioned cookies on all the websites where it is present. \url{criteo.com}, \url{youtube.com}, and \url{creativecdn.com} show very high percentages. Interestingly, Google-owned \url{doubleclick.com}, not in the figure, uses partitioned cookies only on $\approx$12\% of the websites and traditional cookies on $\approx$55\%, highlighting differences between promoted mechanisms and observed deployment patterns.

{Existing, well-established, and effective technologies such as cookies are still preferred to alternatives. Third parties have limited incentives to adopt APIs that potentially limit their data collection capabilities and require engineering effort. CHIPS, being simple, complementary to traditional cookies, and supported by all major browsers, achieves substantial adoption.}

\section{Retired API trend}
\label{sec:deprecated-apis}

\begin{figure}
    \centering
    \begin{subfigure}{\columnwidth}
        \centering
        \includegraphics[width=\linewidth]{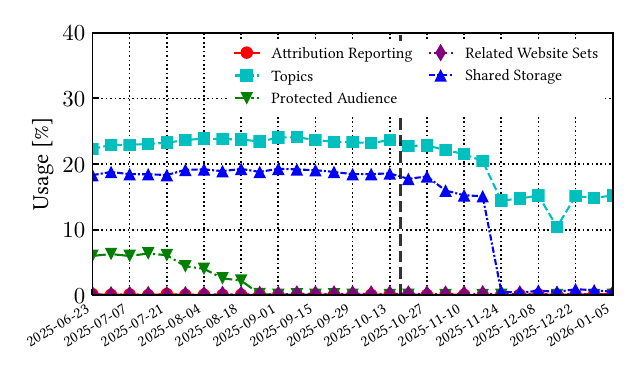}
        \caption{Percentage of websites where a call to a given API is present.}
        \label{fig:deprecated-api-usage}
    \end{subfigure}
    \begin{subfigure}{\columnwidth}
        \centering
        \includegraphics[width=\linewidth]{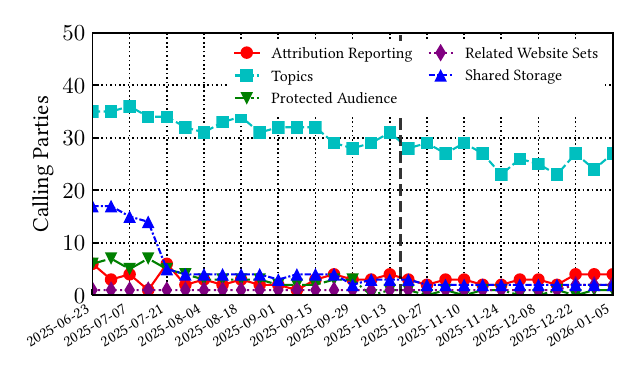}
        \caption{Number of API callers.}
        \label{fig:deprecated-api-callers}
    \end{subfigure}
    \caption{Timeline of retired APIs adoption by third parties. The vertical dashed line highlights the retirement announcement.}
    \label{fig:deprecated-api}
\end{figure}

We now analyse the longitudinal adoption trends of the six Privacy Sandbox APIs that Google announced would be retired, focusing on their deployment by third-party callers over 6 months.
In Figure~\ref{fig:deprecated-api}, we show the per-week measurement timeline. Again, we consider only third-party callers and \afterA. We do not include the Private Aggregation API, given its virtual absence from our dataset. 

Start from the percentage of websites where these APIs are invoked, in Figure~\ref{fig:deprecated-api-usage}. Several observations emerge:
First, the Protected Audience API (green curve) shows the most abrupt decline: its usage drops from around 8\% of websites in late June to nearly zero by early August, well before the API retirement announcement. This contraction aligns with popular adopters, most notably \url{creativecdn.com}, discontinuing their calls during July. After August, the few remaining invocations exclusively initiate auctions, with no evidence of new interest-group assignments. The rapid decline suggests that early adopters discontinued testing or failed to transition from experimentation to sustained deployment.

Second, Topics API and Shared Storage exhibit relatively stable presence until Google’s retirement announcement, after which both APIs show a gradual decline that culminates in a sharp collapse in 24 November 2025 when Shared Storage falls from roughly 19\% of websites to almost zero. Also \url{doubleclick.com} dismissed its usage---see below.

Figure~\ref{fig:deprecated-api-callers} complements this picture by reporting the number of third-party callers for each retired API. 
Across all retired APIs, the number of active callers never exceeds a few dozen and steadily decreases over time, underscoring the highly concentrated nature of Privacy Sandbox experimentation.
The Topics API shows a slow decline that started before the retirement announcement. For Shared Storage, the original 20 callers drop to 3 within a month in July. Yet, the apparent stability in website-level presence despite a sharp reduction in callers is driven by the single large actor \url{doubleclick.com} deploying the API across many sites, masking ecosystem-wide disengagement.

{The same qualitative trends persist when extending the measurements through July 13, 2026 (Appendix~\ref{sec:extended-timeline}), where we observe a stable scenario despite Google's removal plan.}

Overall, our measurements show that the decline of retired Privacy Sandbox APIs was not triggered by {Google's} deprecation announcement, but rather reflects a sustained lack of ecosystem uptake and the withdrawal of {the few} early adopters. The small number of callers, coupled with declining usage well before the announcement, suggests that these APIs failed to transition from experimental deployments to stable production use.

\section{Maintained API trend}
\label{sec:supported-apis}

\begin{figure}
    \begin{subfigure}{\columnwidth}
        \centering
        \includegraphics[width=\linewidth]{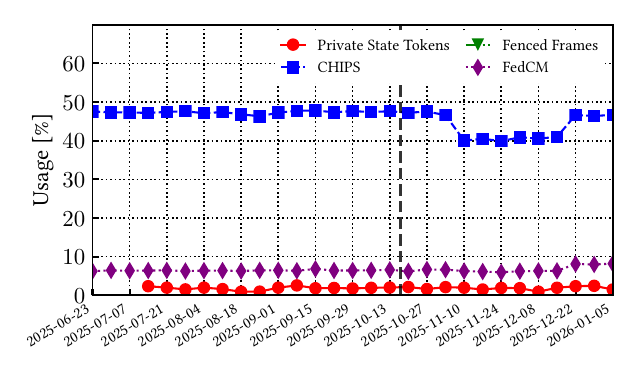}
        \caption{Percentage of websites where a call to a given API is present.}
        \label{fig:supported-api-usage}
    \end{subfigure}
    \begin{subfigure}{\columnwidth}
        \centering
        \includegraphics[width=\linewidth]{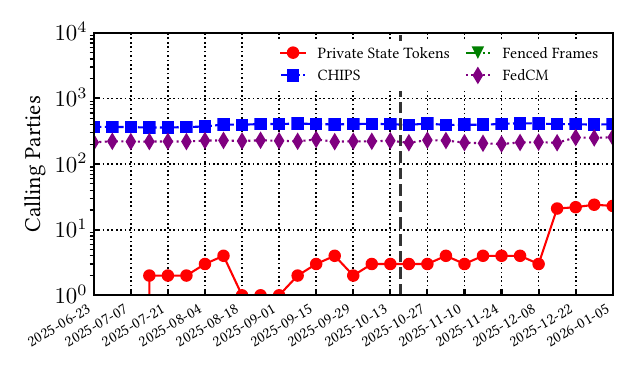}
        \caption{Number of API callers.}
        \label{fig:supported-api-callers}
    \end{subfigure}
    \caption{Timeline of supported API adoption. The vertical dashed line highlights the date of the retirement announcement.}
    \label{fig:supported-api}
\end{figure}

Google opted to maintain four components of the Privacy Sandbox: CHIPS, Private State Tokens, Fenced Frames, and Federated Credential Management. To assess their adoption, we track their usage across our longitudinal measurements, as summarised in Figure~\ref{fig:supported-api}, \afterA case. Here we consider all callers, and not only third parties. Indeed, these APIs do not require the user to complete an attestation process and can be freely used by any service provider.

Figure~\ref{fig:supported-api-usage} shows that, except CHIPS, none of the maintained APIs exceeds the 10\% adoption rate, with no meaningful changes over time. Fenced Frames never appear in our measurements. Their absence from our measurements likely reflects the lack of immediate pressure to replace iframes, rather than measurement limitations.

Figure~\ref{fig:supported-api-callers} summarises the number of callers for each API. Private State Tokens API is invoked by only a handful of players (from 0 to 5 in each snapshot, its variability due to measurement variance), whereas FedCM shows a broader and more diverse caller base. Unlike other APIs, FedCM is primarily invoked by first-party sites, explaining the broader but relatively stable caller base. CHIPS is invoked by a substantially larger and more diverse set of callers (about 350), including many first-party services, consistent with its role as a compatibility-oriented mechanism rather than an advertising-specific API.

Notably, we do not observe any measurable increase in adoption following Google’s announcement that these APIs would remain supported, suggesting that the market has not strongly reacted to Google's strategic signalling.

\begin{figure}
    \centering
    \includegraphics[width=\linewidth]{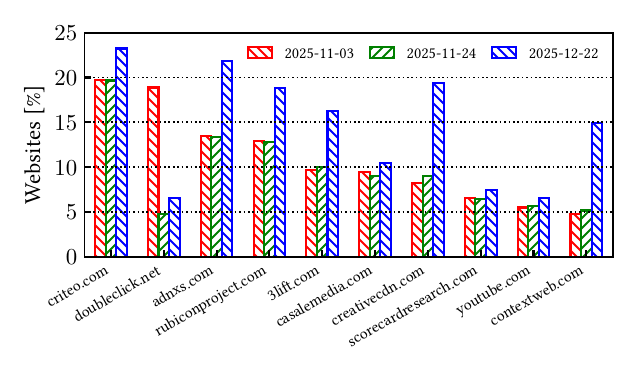}
    \caption{Breakdown of CHIPS usage by third-party caller between 3 November, 24 November, and December 22, 2025.}
    \label{fig:chips-usage-trend}
\end{figure}

We take a closer look at the temporary decrease in the percentage of websites using CHIPS observed between 10 November and 22 December 2025 (Figure~\ref{fig:supported-api-usage}).

Figure~\ref{fig:chips-usage-trend} compares the share of websites where a third party deploys CHIPS across three representative snapshots: before the decrease (3 October 2025), during the decrease (24 November 2025), and after the recovery (22 December 2025). We observe that the decrease and subsequent increase have different underlying dynamics. The initial drop is driven by \url{doubleclick.com}, which substantially reduces the number of websites on which it installs partitioned cookies (only green bar to decrease), while the behaviour of other major third parties remains largely stable.
In contrast, the subsequent increase involves a coordinated rise across multiple third-party callers (all blue bars grow). This pattern suggests a shared implementation change, such as an update in a widely used cookie management library, affecting several actors simultaneously. A more detailed investigation of these mechanisms is left for future work.

{As with the deprecated APIs, we notice no significant changes when extending the measurements to July 2026 (Appendix~\ref{sec:extended-timeline}).}

\section{Questionable usage}
\label{sec:questionable}

\begin{figure}
    \centering
    \begin{subfigure}{\columnwidth}
        \centering
        \includegraphics[width=\linewidth]{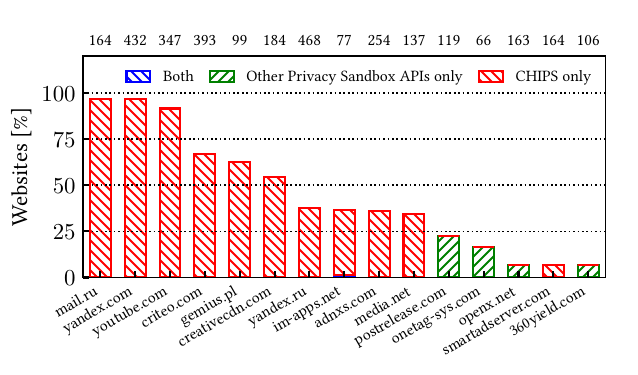}
        \caption{\beforeC}
        \label{fig:chips-vs-others-before}
    \end{subfigure}
    \begin{subfigure}{\columnwidth}
        \centering
        \includegraphics[width=\linewidth]{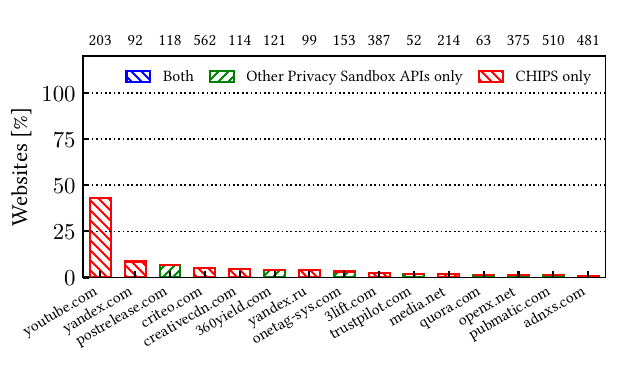}
        \caption{\afterD}
        \label{fig:chips-vs-others-deny}
    \end{subfigure}
    \caption{Comparison between CHIPS and other Privacy Sandbox API usage per top third party during \beforeC and \afterD visits.}
    \label{fig:chips-vs-others}
\end{figure}

We conclude our results by analysing the use of Privacy Sandbox technologies in situations where the user has not yet accepted nor declined the privacy banner (\beforeC), or has explicitly denied consent (\afterD). We refer to both as \textit{questionable} in the sense of being inconsistent with the intended CMP-mediated gating model, without asserting legal non-compliance. 

As previously shown in Figure~\ref{fig:usage-per-api-visit}, we witness a non-negligible fraction of questionable usage.

\subsection{Pervasive third parties}

Figure~\ref{fig:chips-vs-others} reports the third parties responsible for questionable usage. It shows, for each domain, the percentage of websites where it either sets CHIPS cookies only, invokes another privacy-sensitive API, or uses both (similarly to Figure~\ref{fig:chips-vs-others-accept}). All percentages in Figure~\ref{fig:chips-vs-others} are conditional on the third party being present on the site.

Comparing \beforeC and \afterD visits, Figure~\ref{fig:chips-vs-others} shows that the majority of questionable events involve CHIPS only (red pattern). \url{mail.ru}, \url{youtube.com}, and \url{yandex.com} set CHIPS in nearly all websites where they appear, while other actors exhibit more partial or sporadic behaviour.
Since CHIPS are a form of cookie and may include benign, non-privacy-violating cases (e.g., stateful embedded services), we cannot infer malicious intent solely from their presence, and we limit ourselves to documenting this usage.


However, the share of websites showing questionable usage drops sharply after users explicitly deny consent (see Figure~\ref{fig:chips-vs-others-deny}). This indicates that some API-related activity is indeed controlled by consent mechanisms, and that a meaningful part of their usage observed before consent corresponds to components that are  deactivated after user consent denial. For example, \url{mail.ru} sets CHIPS cookies on nearly all the 164 websites where it appears before consent, but largely disappears after consent is denied. Similarly, \url{youtube.com} sets CHIPS cookies on more than 340 websites before consent, while doing so on fewer than 100 websites after consent denial.


A few third parties use other Privacy Sandbox technologies on a fraction of websites where they are present, namely \url{postrelease.com}, \url{360yield.com}, and \url{onetag-sys.com} in both \beforeC and \afterD. Again, this fraction decreases in \afterD, hinting at an undesirable but limited behaviour.
 
Overall, these questionable patterns indicate an immature and inconsistent deployment of Privacy Sandbox technologies rather than deliberate attempts to bypass consent or engage in privacy-violating behaviour. {Consent and compliance mechanisms remain fragile in practice. Any regulatory effort must be supported by continuous auditing and compliance verification mechanisms, to which the research community can contribute.}

\subsection{Questionable CHIPS and traditional cookie usage}

\begin{figure}
    \centering
    \begin{subfigure}{\columnwidth}
        \centering
        \includegraphics[width=\linewidth]{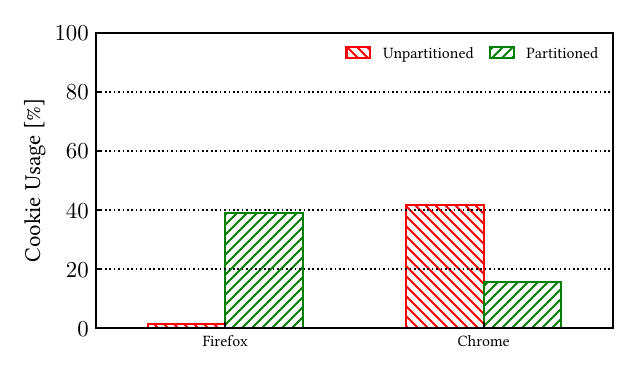}
        \caption{\beforeC}
        \label{fig:cookie-usage-before}
    \end{subfigure}
    \begin{subfigure}{\columnwidth}
        \centering
        \includegraphics[width=\linewidth]{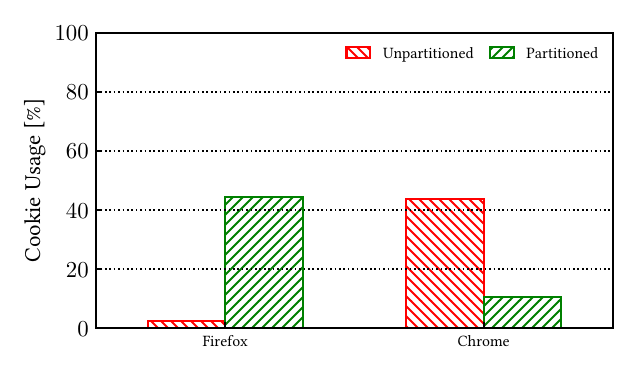}
        \caption{\afterD}
        \label{fig:cookie-usage-deny}
    \end{subfigure}
    \caption{Comparison of persistent third-party cookie usage percentage during \beforeC and \afterD visits.}
    \label{fig:cookie-usage-questionable}
\end{figure}

\begin{figure}
    \centering
    \includegraphics[width=\linewidth]{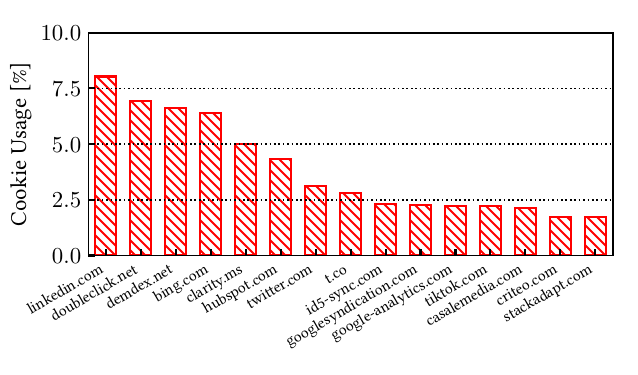}
    \caption{Breakdown of third parties that set an unpartitioned cookie during \afterD on the Chrome browser.}
    \label{fig:questionable-cookie-breakdown}
\end{figure}

Finally, we extend the analysis of Section~\ref{sec:chips-v-trad-cookies-accept}, observing how the usage of unpartitioned and partitioned cookies changes when considering \beforeC and \afterD scenarios. Figure~\ref{fig:cookie-usage-questionable} shows the percentage of websites where we observe a third party installing a cookie, both in Firefox and Chrome. Contrary to Figure~\ref{fig:cookie-usage-after-accept}, here we focus on persistent cookies only, which could possibly be used for tracking. For completeness, we report the cookie duration Cumulative Distribution Function (CDF) in Appendix~\ref{app:duration}.

Compared to \afterA, cookie-setting activity decreases, yet we still observe third-party cookie installation on over 40\% of sites in both \beforeC and \afterD. Firefox forces the usage of partitioned cookies, while Chrome allows for unpartitioned and persistent cookies. Differently from Figure~\ref{fig:chips-vs-others}, we notice no major difference comparing \beforeC and \afterD.

Looking at which are the third-parties setting questionable cookies in \afterD, in Figure~\ref{fig:questionable-cookie-breakdown}, we observe that most of the biggest players in the field are responsible for this behaviour, with \url{linkedin.com}, \url{doubleclick.com}, and \url{demdex.net} leading the way. Observing the same graph (not shown for brevity) for the \beforeC scenario shows similar results, hinting at constant behaviour.

This indicates that third-party state is not consistently gated by the consent flow, motivating deeper auditing of CMP integration and third-party loading logic as seen in~\cite{verna2025understanding}. Investigating possible legal implications is outside the scope of this paper and left for future work.

\section{Conclusions}
\label{sec:conclusion}

In this paper, we analysed the final months of Google's Privacy Sandbox initiative, measuring how these APIs were deployed across the Web during the period preceding their retirement. Our longitudinal results show that the deprecation announcement merely formalised an ongoing trend: most APIs saw limited adoption and were already being abandoned by the ecosystem, with only a small set of actors ever testing them.


Only CHIPS, i.e., partitioned cookies, achieved broad and sustained usage. However, in Google Chrome, their adoption has not come at the expense of classical, unpartitioned third-party cookies, which continue to be by far the most popular---as well as the most privacy-invasive---means to manage stateful services. 

All other APIs are approaching irrelevance. Protected Audience and Shared Storage usage declined to near-zero months before the API retirement announcement, and Topics API is following a similar trajectory. For the APIs that Google will maintain, we observe no momentum: Private State Tokens, Fenced Frames, and FedCM appear on only a small fraction of sites, with no sign of growth.

{The technological push is insufficient to stimulate the adoption of a new technology. On their own, a sound, although debatable, solution and even the support of major players like Google were insufficient to drive ecosystem-wide adoption. We conjecture that only a regulatory framework could enforce a major shift in the privacy-preserving ads landscape. Therefore, the effective deployment of future proposals requires stronger and up-to-date regulatory pressure.}

A framework originally presented as a compromise between privacy and advertising ultimately failed to gain traction, and Google finally abandoned it. With the Privacy Sandbox fading without leaving a viable alternative, the challenge of enabling interest-based yet privacy-preserving advertising remains open, calling for renewed efforts to design privacy-preserving advertising mechanisms that better align deployability, incentives, and enforcement.

\begin{acks}
    This work has received funding from the Applied Sciences Italian Fund (Fondo Italiano per le Scienze Applicate---FISA) by the Italian Ministry of University and Research, under the AI4CTI project (grant agreement No. FISA-2023-00168).
\end{acks}

\bibliographystyle{ACM-Reference-Format}
\bibliography{refs}

\appendix

\section{Alternative Visualisations}
\label{sec:alternative-visualisations}

\subsection{CP-focused flow diagram}
\label{sec:cp-focused-flow}

\begin{figure}
    \centering
    \includegraphics[width=\linewidth]{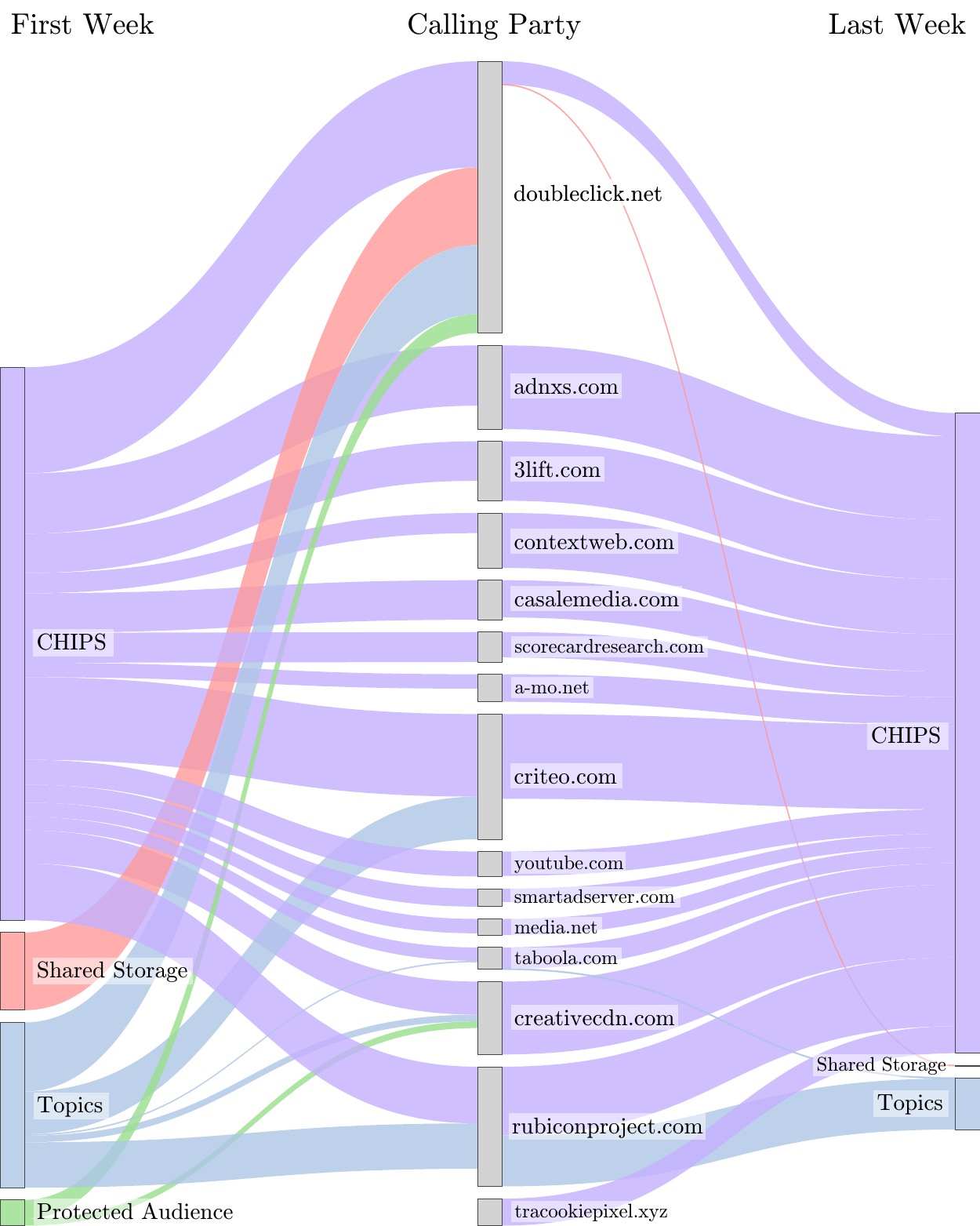}
    \caption{Caller-centric flow graph of third-party callers during the latest campaign.}
    \label{fig:third-party-callers-alt}
\end{figure}

Figure~\ref{fig:third-party-callers-alt} offers an alternative view of Figure~\ref{fig:third-party-callers}, which focuses on the difference between the callers' presence rather than the APIs called.

\subsection{Extended Usage Timelines}
\label{sec:extended-timeline}

\begin{figure}
    \centering
    \begin{subfigure}{\columnwidth}
        \centering
        \includegraphics[width=\linewidth]{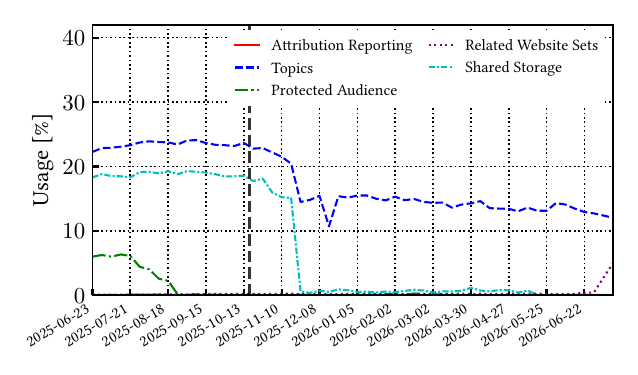}
        \caption{Percentage of websites where a call to a given API is present.}
        \label{fig:deprecated-api-usage}
    \end{subfigure}
    \begin{subfigure}{\columnwidth}
        \centering
        \includegraphics[width=\linewidth]{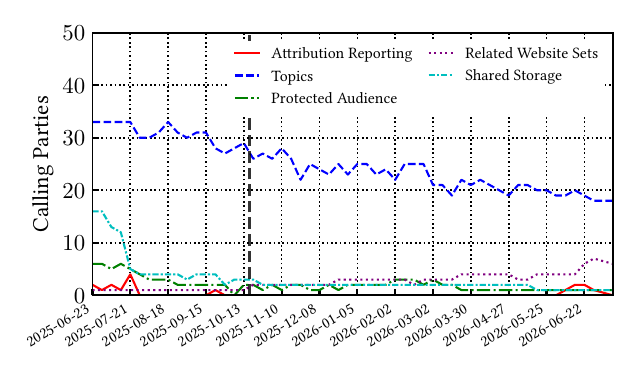}
        \caption{Number of API callers.}
        \label{fig:deprecated-api-callers}
    \end{subfigure}
    \caption{Extended timeline of retired APIs adoption by third parties. The vertical dashed line highlights the retirement announcement.}
    \label{fig:extended-deprecated-api}
\end{figure}

\begin{figure}
    \begin{subfigure}{\columnwidth}
        \centering
        \includegraphics[width=\linewidth]{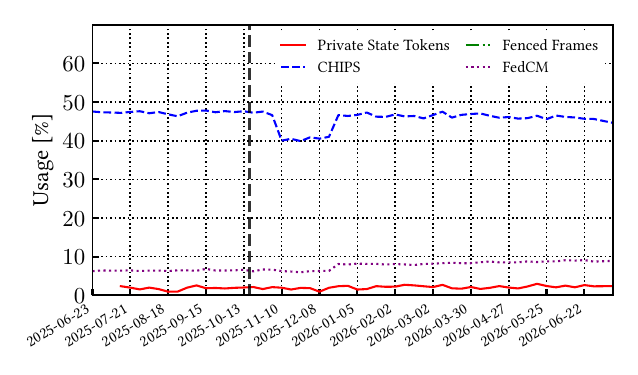}
        \caption{Percentage of websites where a call to a given API is present.}
        \label{fig:supported-api-usage}
    \end{subfigure}
    \begin{subfigure}{\columnwidth}
        \centering
        \includegraphics[width=\linewidth]{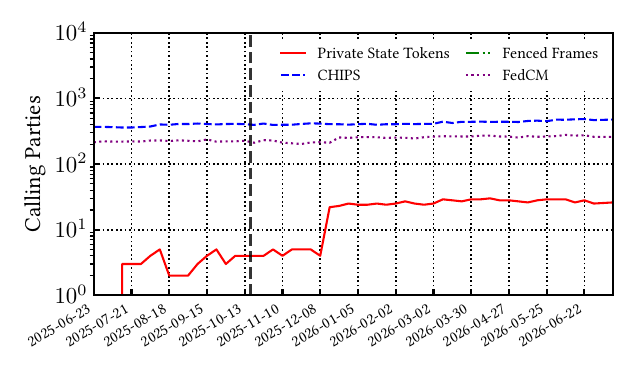}
        \caption{Number of API callers.}
        \label{fig:supported-api-callers}
    \end{subfigure}
    \caption{Extended timeline of supported API adoption. The vertical dashed line highlights the date of the retirement announcement.}
    \label{fig:extended-supported-api}
\end{figure}

In this appendix, we report extended versions of the API usage timelines through July 13, 2026. Figures~\ref{fig:extended-deprecated-api} and~\ref{fig:extended-supported-api} extend Figures~\ref{fig:deprecated-api} and~\ref{fig:supported-api}, respectively, reporting both API usage and the number of calling parties over the extended measurement period.

\section{Cookie duration}
\label{app:duration}

\begin{figure}
    \centering
    \begin{subfigure}{\columnwidth}
        \centering
        \includegraphics[width=\linewidth]{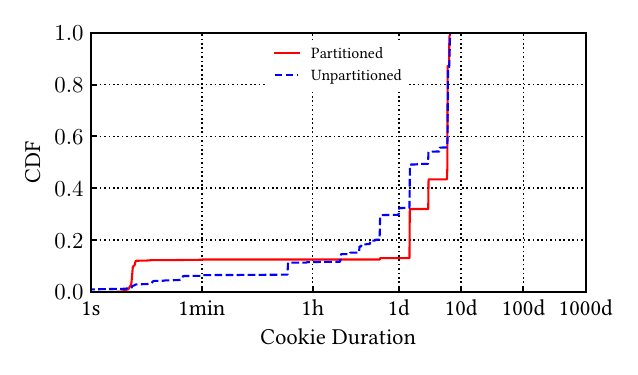}
        \caption{\afterA}
        \label{fig:cookie-duration-accept}
    \end{subfigure}
    \begin{subfigure}{\columnwidth}
        \centering
        \includegraphics[width=\linewidth]{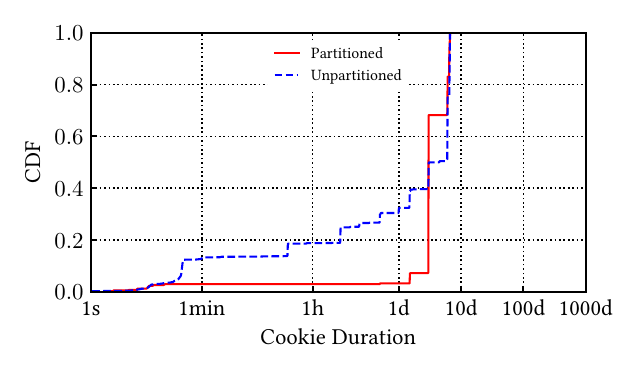}
        \caption{\afterD}
        \label{fig:cookie-duration-deny}
    \end{subfigure}
    \caption{Distribution of persistent cookie duration in \afterA and \afterD.}
    \label{fig:cookie-duration}
\end{figure}

Figure~\ref{fig:cookie-duration} compares the cookie duration in \afterA and \afterD, for both partitioned and unpartitioned cookies. Most cookies have a duration of several days, even in the \afterD visit. This strengthens their questionable usage.

\section{Ethical considerations}
This work did not present severe ethical aspects to consider. We ran a crawling campaign to some of the most visited websites in the world, issuing up to six web page visits once a week. We deem this a negligible volume of traffic, which we expect did not cause any problem to websites of such size.

\section{Open Science}
\label{app:ethics}
This work is based on the results of an extensive, specialised Web crawling campaign and subsequent data analysis. To enable independent verification and facilitate reproduction of our results, we release the code and data we produced in the following Git repository: \url{https://github.com/AI4CTI/priv-accept-ps}. 

The repository is organised into the following components:
\begin{itemize}
    \item The \path{crawler/} folder contains the Python-based crawler used to visit individual websites and collect measurement data in JSON format;
    \item The \path{post-process/} folder contains the code used to filter and pre-process the raw Web logs produced by the crawler;
    \item The \path{extract-allowed/} folder contains scripts that extract the set of domains that have enrolled to use certain Privacy Sandbox components;
    \item The \path{open-data/} folder contains the filtered data, plus the code that analyses it to produce the plots with the same procedure we employed in the paper.
\end{itemize}
Each component has its own \path{README.md} file containing additional information.

To support reproducibility of the measurement workflow, we provide a set of scripts that automate the setup and execution of the crawling and analysis pipeline:
\begin{itemize}
    \item \path{install-dependencies.sh}, which installs the required software dependencies;
    \item \path{build.sh}, which builds the repository’s components as Docker containers;
    \item \path{analyze-ps-complete.sh}, which executes the complete measurement campaign over a configurable set of $N$ websites and produces the final analysis outputs as zipped CSV files.
\end{itemize}
These scripts were verified to work on a fresh installation of Ubuntu~22.04~LTS.

\section{AI Assistance Disclosure}
The authors used generative AI-based tools to revise the text, improve flow and correct any typos, grammatical errors, and awkward phrasing.
The authors are fully responsible for the content of the paper, including the study design, data collection, analysis, interpretation of results, and all conclusions.

\end{document}